\documentclass[reprint,amsmath,amssymb,aps,prb,showpacs]{revtex4-1}

\usepackage{graphicx}
\usepackage{amsmath, amssymb}

\usepackage[normalem]{ulem} 

\renewcommand{\Re}{\mathop{\mathrm{Re}}\nolimits}
\renewcommand{\Im}{\mathop{\mathrm{Im}}\nolimits}

\begin{document}
\title{$\omega$--$k_x$ Fano line shape in photonic crystal slabs}

\author{Dmitry A. Bykov}
 \email{bykovd@gmail.com}
\author{Leonid L. Doskolovich}%
 \email{leonid@smr.ru}
\affiliation{Image Processing Systems Institute of the RAS, 151 Molodogvardeiskaya st., Samara, 443001, Russia}
\affiliation{Samara State Aerospace University (SSAU), 34 Moskovskoye shosse, Samara, 443086, Russia}

\date{\today}



\begin{abstract}
We study the resonant properties of photonic crystal slabs theoretically.
An $\omega$--$k_x$ Fano line-shape that approximates the transmission (reflection) spectrum is obtained.
This approximation, being a function of light's frequency and in-plane wave vector, generalizes the conventional Fano line-shape.
Two particular approximations, parabolic and hyperbolic, are obtained and investigated in detail, taking into account the symmetry of the structure, the reciprocity, and the energy conservation.
The parabolic approximation considers a single resonance at normal incidence, while the hyperbolic one takes into account two modes, the symmetric and the antisymmetric. 
Using rigorous simulations based on the Fourier modal method we show that the hyperbolic line-shape provides a better approximation of the transmission spectrum.
By deriving the causality conditions for both approximations, we show that only the hyperbolic one provides causality in a relativistic sense.
\end{abstract}

\pacs{42.70.Qs, 78.67.Pt, 42.79.Dj}
\maketitle

\section{\label{section1}Introduction}
Fano resonances have attracted much attention in the past few decades.
While originally developed to describe the atomic absorption spectrum~\cite{Beutler:1935:zp, Fano:1935:nc} and the electron energy loss spectrum~\cite{Fano:1961:pr}, they later found numerous applications in different areas of physics including solid state physics and optics~\cite{Miroshnichenko:2010:rmp, Collin:2014:rpp, Zhou:2014:pqe}.

Fano resonance occurs when two scattering processes take place simultaneously: the resonant scattering and non-resonant one. 
The interaction between these two processes results in a distinctive asymmetric line-shape of the scattering amplitude, which is called the Fano line-shape. 
This model for resonant scattering allows one to write the following simple approximation for the scattering amplitude as a function of frequency: 
\begin{equation}
\label{Fano}
T(\omega) \approx t + \frac{s}{\omega - \omega_p} = t \frac{\omega - \omega_z}{\omega - \omega_p}.
\end{equation}
Here, $t$ is the non-resonant scattering coefficient, $\omega_p$ and $\omega_z$ are the pole and the zero of the function $T(\omega)$.
Equation~\eqref{Fano} is widely used to explain resonant phenomena in the transmission and reflection spectra of optical resonators~\cite{Kirilenko:1993:em, Centeno:2000:prb, Miroshnichenko:2010:rmp},
diffraction gratings and photonic crystal slabs~\cite{Collin:2001:prb, Tikhodeev:2002:prb, Sarrazin:2003:prb, Gippius:2005:prb, Lomakin:2006:trant, my:bykov:2013:jlt, Zhou:2014:pqe, Collin:2014:rpp}.
By replacing the frequency $\omega$ in~\eqref{Fano} with the in-plane wave vector component $k_x$, one can investigate the \emph{angular} spectrum of resonant diffractive structures~\cite{Popov:1986:oa, Neviere:1995:josaa, Lomakin:2007:trant}. 
Moreover, the resonant approximations of the scattering amplitude as a function of both $\omega$ and $k_x$ were proposed~\cite{Shipman:2005:pre, Shipman:2013:jmp, Lalanne:2008:nat}.

Of particular interest is studying the Fano resonances in symmetric structures made of lossless reciprocal materials. 
For these structures, a special form of the Fano line-shape~\eqref{Fano} can be obtained, revealing a number of intriguing optical effects such as the total transmission and total reflection of the incident light~\cite{Popov:1986:oa, Fan:2003:josaa, Gippius:2005:prb, Miroshnichenko:2010:rmp}.

\begin{figure}[hbt]
	\includegraphics{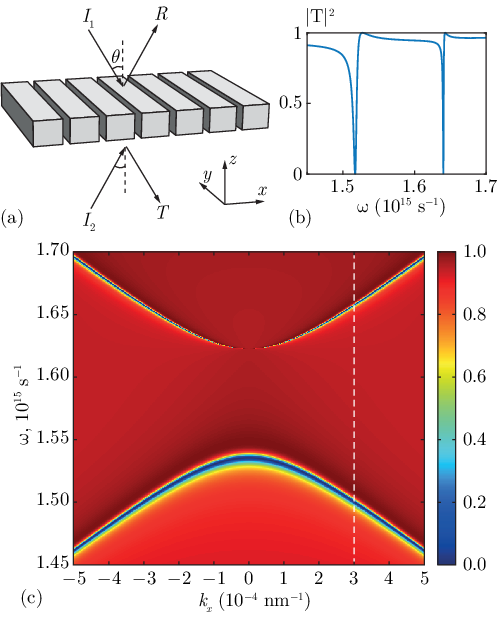}
	\centering
  \caption{\label{fig1}(Color online) (a)~Geometry of 1D PCS (parameters: period $d=1000\,\rm{nm}$; height $h=700\,\rm{nm}$; fill-factor $4/5$; surrounding medium refractive index $n_{\rm s} = 1$; structure material permittivity $\varepsilon_{\rm gr} = 2$).
	The structure is invariant in the $y$-direction. (b)~Transmission coefficient $|T|^2$ vs. the incident light's angular frequency $\omega$ at fixed wave vector in-plane component $k_x=3\times10^{-4}\,{\rm nm}^{-1}$ for TM-polarized incident wave.
	(c)~Transmission coefficient $|T|^2$ vs. the incident light's angular frequency $\omega$ and the wave vector in-plane component $k_x=(\omega/c) n_{\rm s} \sin \theta$ for TM-polarized incident wave.}
\end{figure}

In this paper, we study the resonances of 1D photonic crystal slabs [Fig.~\ref{fig1}(a)].
Photonic crystal slabs (PCS) or diffraction gratings are planar optical structures that are periodic in one or two transverse directions.
Such structures exhibit resonant features in the transmission and reflection spectra, which were studied for the first time by R.~W.~Wood in 1902~\cite{Wood:1902:pm}.
These features correspond to the Fano resonances and are explained in terms of the excitation of the quasiguided modes (either of plasmonic nature or not)~\cite{Hessel:1965:ao}. 
Due to the pronounced resonant features, the PCS are widely used as optical filters~\cite{Wang:1993:ao, Zhou:2014:pqe}, polarizers~\cite{Mutlu:2012:ol}, lasers and sensors~\cite{Zhou:2014:pqe}.
Other applications include beam and pulse shaping~\cite{Lomakin:2006:trant, Lomakin:2007:trant, my:golovastikov:2013:jopt}, enhancing nonlinear and magneto-optical effects~\cite{Belotelov:2014:prb, Liu:2001:prb, Neviere:1995:josaa, Zhou:2014:pqe}, controlling optical properties mechanically~\cite{Zhou:2014:pqe}.

A typical $\omega$--$k_x$ transmission spectrum of the PCS rigorously calculated using the Fourier modal method~\cite{Moharam:1995:josaa, Li:1996:josaa} is presented in Fig.~\ref{fig1}(c). 
Similar spectra were observed both theoretically and experimentally in a number of papers~\cite{Christ:2003:prl, Yablonskii:2002:pssa, Belotelov:2014:prb}.
The computed spectrum in Fig.~\ref{fig1}(c) demonstrates resonant minima governed by the excitation of quasiguided modes.
According to Fig.~\ref{fig1}(b) the resonances have pronounced asymmetric Fano line-shape.
The points of minimal transmission form two branches separated by a bandgap in the center of the first Brillouin zone.
Let us note that the known $\omega$--$k_x$ approximations proposed in Refs.~[\onlinecite{Shipman:2005:pre, Shipman:2013:jmp}] have a limited applicability and allow one to describe only one branch of the transmission coefficient. The model proposed in Ref.~[\onlinecite{Lalanne:2008:nat}] takes account of two branches, but it describes only perforated metal films with extraordinary optical transmission.

In this paper, we derive new $\omega$--$k_x$ approximations for the complex transmission (reflection) spectrum [Fig.~\ref{fig1}(c)] of the PCS in the vicinity of resonances.
The proposed approximations generalize the conventional Fano line-shape~\eqref{Fano} of one argument. 
For the first time, the $\omega$--$k_x$ Fano line-shapes are obtained taking into account the structure symmetry and reciprocity, the energy conservation law, and the causality condition. 
We believe that the results of the current paper are important for design of a wide range of photonic devices, such as guided-mode resonant filters, resonant structures for spatiotemporal pulse shaping, sensors, lasers, magneto-optical and non-linear devices. 
In particular, the $\omega$--$k_x$ Fano line-shape can be used to describe a general class of spatiotemporal transformations of optical beams implemented by resonant diffraction structures.

The paper is organized in five sections. 
Following the Introduction, Section~\ref{section2} presents the rigorous derivation of the $\omega$--$k_x$ Fano line-shape.
Two types of approximation, parabolic and hyperbolic, are obtained.
In the following sections we give physical interpretation of the approximations' parameters and determine the relations between them. 
In particular, the consequences of the structure symmetry are studied in Section~\ref{section3} while the causality condition is discussed in Section~\ref{section4}.

\section{\label{section2}The \lowercase{$\omega$--$k_x$} Fano line-shape}
Consider a 1D periodic structure (PCS or diffraction grating) with period $d$ [Fig.~\ref{fig1}(a)].
In this section, we derive the approximations for the transmission (reflection) coefficient that generalizes the well-known Fano line-shape~\eqref{Fano}.
We assume that the structure is subwavelength and supports only zeroth propagating diffraction orders.
For the sake of simplicity, in this section we assume that the structure has two symmetry planes: $xOy$ and $yOz$ (other symmetries will be discussed in Section~\ref{section3}).

Consider two monochromatic plane waves incident on the structure at the angle $\theta$ from the superstrate and substrate regions [Fig.~\ref{fig1}(a)]. 
In this case, the reflection and transmission can be described by the scattering matrix~$S$ which relates the amplitudes of the incident waves and of the scattered waves (zeroth diffraction orders): 
\begin{equation}
\label{smatrix}
\begin{bmatrix}
R\\T
\end{bmatrix}
=
S
\begin{bmatrix}
I_1\\I_2
\end{bmatrix}
,
\end{equation}
where $T$ and $R$ are the complex amplitudes of the scattered  waves; 
$I_1,I_2$ are the complex amplitudes of the incident waves. 
For specified PCS and fixed polarization the scattering matrix is a function of the incident light angular frequency $\omega$ and of the in-plane wave vector component $k_x=k_0 n_{\rm s} \sin\theta$, 
where $n_{\rm s}$ is the surrounding medium  refractive index, and $k_0=\omega/c$ is the wavenumber. 

The modes of the structure are the field distributions that exist in the structure in the absence of the incident light (at $I_1 = I_2 = 0$).
According to Eq.~\eqref{smatrix}, the modes of the PCS are defined by the following homogeneous system of linear equations:
\begin{equation}
\label{smatrixmode}
S^{-1}
\begin{bmatrix}
R\\T
\end{bmatrix}
=
0.
\end{equation}
Nontrivial solution of this system are given by the equation $\det S^{-1}(k_x, \omega) = 0$.
By denoting $l(k_x,\omega) = \det S^{-1}(k_x, \omega)$, we can write the dispersion equation of the modes of the structure in the following form: 
\begin{equation}
\label{dispeq}
l(k_x, \omega) = 0.
\end{equation}

In this paper, we are interested in the (quasiguided) modes that can be excited by the incident plane wave. 
Due to reciprocity condition, these modes will scatter away from the structure.
This means that the mode amplitudes will decay in time.
Hence, the mode frequencies are the complex numbers with a negative imaginary part (for $\mathrm{e}^{-\mathrm{i} \omega t}$ time convention).
Indeed, the dispersion equation~\eqref{dispeq} is usually solved for complex angular frequency~$\omega$ at real $k_x$~\cite{Tikhodeev:2002:prb, Fan:2003:josaa, my:bykov:2013:jlt}.
In this case, $\omega$ satisfying Eq.~\eqref{dispeq} is the complex pole of the scattering matrix~$S$~\cite{my:bykov:2013:jlt} and, consequently, of the functions~$T$ and~$R$.
On the other hand, one can define real $\omega$ and solve Eq.~\eqref{dispeq} for the complex $k_x$~\cite{Neviere:1995:josaa}.
Note that in order to work with complex frequencies and/or complex wavenumbers one should formally replace $\det S^{-1}(k_x, \omega)$ with its analytical continuation~\cite{Tikhodeev:2002:prb}; in what follows we suppose that the function $l(k_x, \omega)$ depends analytically on $\omega$ and $k_x$.


Following the approach used in Ref.~[\onlinecite{Shipman:2005:pre}], let us apply the Weierstra{\ss} preparation theorem~\cite{Scheidemann:2005} to the function $l(k_x, \omega)$.
To do this we suppose that the resonance of the PCS at $k_x = 0$ corresponds to the mode with the complex frequency $\omega_p$ [i.e.\ $l(0,\omega_p) = 0$]. 
Moreover, since we consider the structure with the vertical symmetry plane, $T(k_x, \omega) = T(-k_x, \omega)$. Therefore, we obtain $\partial T/\partial k_x|_{k_x = 0} = 0$ and, consequently, $\partial l/\partial k_x|_{k_x = 0} = 0$. 
In this case, the Weierstra{\ss} preparation theorem~\cite{Scheidemann:2005} gives the following representation of $l(k_x, \omega)$:
\begin{equation}
\label{w}
	l(k_x, \omega) = \left[k_x^2 + A(\omega)\right] \Phi(k_x, \omega),
\end{equation}
where $A(\omega_p) = 0$ and $\Phi(k_x, \omega)$ is an analytic function which is non-zero in a vicinity of $(0, \omega_p)$.
Let us note that as distinct from Ref.~[\onlinecite{Shipman:2005:pre}], we used the Weierstra{\ss} preparation theorem for the $k_x$-variable (rather than $\omega$). This will further allow us to obtain an approximation of the PCS transmission coefficient taking account of two branches in Fig.~\ref{fig1}(c).

Let us now consider a plane wave of unit amplitude that is incident on the structure from the superstrate region ($I_1 = 1, I_2 = 0$). 
In this case,~$R$ and~$T$ will correspond to the complex reflection and transmission coefficients.
The latter can be obtained by solving Eq.~\eqref{smatrix} using Cramer's rule:
\begin{equation}
\label{Eq4}
T = \frac{\det Q}{\det S^{-1}},
\end{equation}
where $Q$ is the matrix formed by replacing the second column of $S^{-1}$ by the column vector $[1\;\;0]^{\rm T}$.
By substituting Eq.~\eqref{w} into the last equation, we obtain
\begin{equation}
\label{Eq5}
T(k_x,\omega) = \frac{\det Q(k_x,\omega)/\Phi (k_x,\omega)}{k_x^2 + A(\omega)}.
\end{equation}

By replacing the numerator and the denominator in Eq.~\eqref{Eq5} with their Taylor polynomials of different degrees, we can obtain different resonant approximations for the transmission coefficient.

\subsection{Parabolic approximation}
Let us expand the function $A(\omega)$ in Eq.~\eqref{Eq5} in Taylor series at $\omega = \omega_p$ up to the linear term:
\begin{equation}
\label{Eq6}
T(k_x,\omega) = \frac{\det Q(k_x,\omega)/\Phi (k_x,\omega)}{k_x^2 - \beta(\omega - \omega_p)}.
\end{equation}
According to the Weierstra{\ss} preparation theorem, the function $1/\Phi(k_x, \omega)$ is analytic in a vicinity of $(0, \omega_p)$ and hence we can expand the numerator in Eq.~\eqref{Eq6} in a Taylor series around this point:
\begin{equation}
\label{parabolic2}
T(k_x,\omega) 
= t\frac{k_x^2 - \alpha(\omega-\omega_p) + \alpha_2}{k_x^2 - \beta(\omega-\omega_p)}
= t\frac{k_x^2 - \alpha(\omega-\omega_z)}{k_x^2 - \beta(\omega-\omega_p)}.
\end{equation}
Here $\omega_z$ and $\omega_p$ are the pole and the zero of the transmission coefficient at normal incidence of light (at $k_x = 0$).
Let us note that we expanded the numerator up to the terms of the same order as the denominator.
In this case, the $T(k_x, \omega)$ is a bounded function for large values of arguments. 
According to (1), we will refer to $t$ as the non-resonant transmission coefficient.
The following form of Eq.~\eqref{parabolic2} will be useful for the subsequent analysis:
\begin{equation}
\label{parabolic}
T(k_x,\omega) = t\frac{k_x^2 + z_0 + z_1 \omega}{k_x^2 + p_0 + p_1 \omega}.
\end{equation}
If we equate to zero the numerator of the last expression, we obtain the equation of parabola. 
By equating to zero the denominator of Eq.~\eqref{parabolic} we obtain the mode dispersion equation that defines a parabola as well.
This is the reason that we refer to the representations~\eqref{parabolic2},~\eqref{parabolic} as \emph{parabolic} approximations.

Note that at fixed~$k_x$ Eq.~\eqref{parabolic} will coincide with the known Fano line-shape~\eqref{Fano}.
At the same time, at fixed~$\omega$ Eq.~\eqref{parabolic} describes the Fano line-shape as a function of~$k_x^2$.
This means that Eqs.~\eqref{parabolic2},~\eqref{parabolic} are the generalizations of Eq.~\eqref{Fano}.
In what follows, we will refer to them as the $\omega$--$k_x$ Fano line-shape.

\begin{figure*}
	\includegraphics{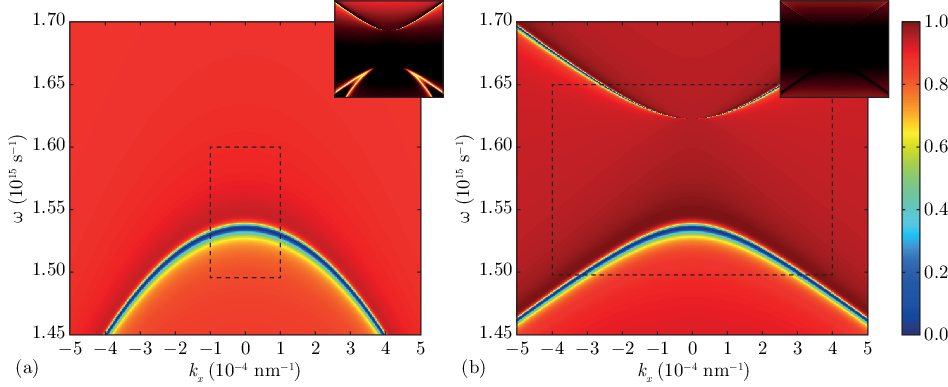}
	\centering
		\caption{\label{fig2}(Color online) Approximated transmission coefficient: (a)~parabolic approximation, (b)~hyperbolic approximation.
		The insets represent the squared modulus of the approximation error. 
		The latter inside the indicated rectangular areas does not exceed~$0.1$.}
\end{figure*}

We used Eq.~\eqref{parabolic2} to approximate the transmission coefficient of the PCS (Fig.~\ref{fig1}).
To do this, we calculated the pole $\omega_p = 1.5361\times 10^{15}-3.195 \times 10^{12}\mathrm{i} \,\rm{s}^{-1}$ using the scattering matrix approach~\cite{Tikhodeev:2002:prb, my:bykov:2013:jlt}.
The transmission zero $\omega_z = 1.5352\times 10^{15} \,\rm{s}^{-1}$ was obtained by finding a minimum of the rigorously calculated transmission coefficient $T(\omega)$ at $k_x = 0$.
To calculate $T(\omega)$, we used the Fourier modal method~\cite{Moharam:1995:josaa, Li:1996:josaa}.
The remaining parameters ($\alpha = -5.4726\times 10^{20} \,\rm{s}/\rm{nm}^2$, $\beta = -5.4809\times10^{20}+3.0813\times 10^{18}\mathrm{i} \,\rm{s}/\rm{nm}^2$, and $t = 0.0186-0.9314\mathrm{i}$) were determined by the least squares fitting of the rigorously calculated transmission coefficient in Fig.~\ref{fig1}(c).
The result of the approximation is shown in Fig.~\ref{fig2}(a).
One can see that Eq.~\eqref{parabolic2} approximates only one branch of the transmission coefficient in a vicinity of the point ($k_x = 0, \omega = \Re\omega_p$).
Moreover, the parabola does not describe the mode dispersion at the large values of $k_x$.
In order to overcome these issues let us construct a higher-order approximation.

\subsection{Hyperbolic approximation}
Let us expand the function $A(\omega)$ in Eq.~\eqref{Eq5} in a Taylor series at $\omega=\omega_p$ up to the second-order term.
Then we replace the numerator with its Taylor polynomial of the same degree as the denominator.
As a result, we obtain the following \emph{hyperbolic} approximation of the transmission coefficient: 
\begin{equation}
\label{hyperbolic}
T(k_x,\omega) = t\frac{k_x^2 + z_0 + z_1 \omega + z_2 \omega^2}{k_x^2 + p_0 + p_1 \omega + p_2 \omega^2}.
\end{equation}
By factoring the $\omega$-polynomials we rewrite Eq.~\eqref{hyperbolic} in the following form: 
\begin{equation}
\label{hyperbolic2}
T(k_x,\omega) = t\frac{v_g^2 k_x^2 - \gamma(\omega-\omega_{z1})(\omega-\omega_{z2})}{v_g^2 k_x^2 - (\omega-\omega_{p1})(\omega-\omega_{p2})},
\end{equation}
where $v_g^2 = -1/p_2$, $\gamma = z_2/p_2$. 
Note that $\omega_{z1}$ and $\omega_{z2}$ are the zeros of the transmission coefficient at normal incidence (at $k_x = 0$),
while $\omega_{p1}$ and $\omega_{p2}$ are the poles of the transmission coefficient at normal incidence.

Figure~\ref{fig2}(b) shows the transmission coefficient approximation calculated using Eq.~\eqref{hyperbolic2}. 
In order to obtain this approximation we calculated the second pole--zero pair ($\omega_{p2} = \omega_{z2} = 1.6222\times 10^{15} \,\rm{s}^{-1}$).
Let us note that the calculated second pole and zero are equal to each other; we discuss this fact later in Section~\ref{section3}B.
The remaining parameters ($v_g = 2.1834\times10^{17}\,{\rm nm}/{\rm s}$, $\gamma = 1.0014-0.00518\mathrm{i}$, $t = 0.1571 - 0.9520\mathrm{i}$) were estimated by means of optimization. 
According to Fig.~\ref{fig2}(b), Eq.~\eqref{hyperbolic2} approximates the transmission coefficient much better in comparison with Eq.~\eqref{parabolic2}. 
The reason is that the hyperbolic approximation takes account of two poles at normal incidence. 
Besides, the mode dispersion law is better described with a hyperbola rather than with a parabola.

The approximations \eqref{parabolic2}--\eqref{hyperbolic2} are obtained for the transmission coefficient. 
However, similar equations can also be used for the reflection coefficient. 
Moreover, according to Eqs.~\eqref{smatrixmode} and~\eqref{Eq4}, the denominators for reflection and transmission approximations coincide, while the numerators are generally different. 

Following the parabolic and hyperbolic approximations we can obtain even higher-order approximations by expanding the function in Eq.~\eqref{Eq5} up to the terms of higher order.
These approximations can be used to approximate the transmission spectrum in a broader $\omega$-range.

In this section, we assumed that the transmission (reflection) coefficient is an even function of $k_x$.
This is true for the symmetric structure shown in Fig.~\ref{fig1}(a). 
The validity of this assumption for the structures of different symmetries is discussed in the following section.

\section{\label{section3}Symmetry, reciprocity and energy conservation}
\begin{figure}[hbt]
	\includegraphics{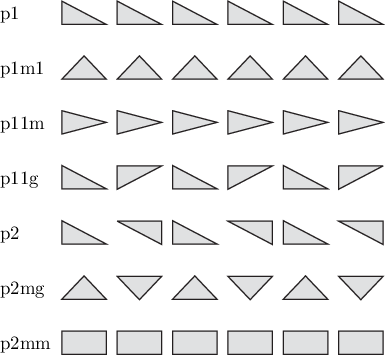}
	\centering
		\caption{\label{fig3}Photonic crystal slabs with different symmetries: seven frieze symmetry groups.}
\end{figure}

The symmetry of 1D PCS can be described by one of seven frieze symmetry groups~\cite{Kopsky:2002}.
Photonic crystal slabs of different symmetries are shown in Fig.~\ref{fig3}. 
In this section, we will discuss the most important of these symmetries and their consequences. 
In particular, we will consider special forms of approximations~\eqref{parabolic}~and~\eqref{hyperbolic} that take into account the symmetries of the structure.

\subsection{Symmetry and reciprocity}
Equations~\eqref{w},~\eqref{parabolic2}--\eqref{hyperbolic2} were derived assuming that the scattering amplitude (transmission or reflection coefficient) is an even function of $k_x$.
For reflection, this is true due to the reciprocity~\cite{Popov:1986:oa, Gippius:2005:prb}. 
For transmission, however, some assumptions on the symmetry of the structure should be made in order to provide $T(k_x,\omega) = T(-k_x,\omega)$.

To study the symmetry of the structure let us consider the elements of the scattering matrix as functions of $k_x$.
The reciprocity and symmetry conditions impose restrictions on the scattering matrix form~\cite{Popov:1986:oa, Gippius:2005:prb}.
In Table~\ref{table1} we present the general form of the scattering matrix for each of seven frieze symmetry groups.
In this table the even functions have $k_x^2$ as the argument.

\begin{table}[tb]
	\caption{\label{table1}General form of the scattering matrix for the structures of different symmetry.}
	\begin{ruledtabular}
		\begin{tabular}{c c}
			Symmetry group & \hspace{1em} Scattering matrix $S$ \hspace{1em}
			 \\ \hline
			p1 & \rule{0pt}{4.5ex} $\begin{bmatrix}R_1(k_x^2)& T(k_x)\\ T(-k_x) &R_2(k_x^2)\end{bmatrix}$  \vspace{0.2em}
			\\
			p2 & $\begin{bmatrix}R(k_x^2)& T(k_x)\\ T(-k_x) &R(k_x^2)\end{bmatrix}$  \vspace{0.2em}
			\\
			p1m1 & $\begin{bmatrix}R_1(k_x^2)& T(k_x^2)\\ T(k_x^2) &R_2(k_x^2)\end{bmatrix}$  \vspace{0.2em}
			\\
			 \hspace{1em} p11m, p11g, p2mg, p2mm \hspace{1em} & $\begin{bmatrix}R(k_x^2)& T(k_x^2)\\ T(k_x^2) &R(k_x^2)\end{bmatrix}$  \vspace{0.2em}
		\end{tabular}
	\end{ruledtabular}
\end{table}

According to Table~\ref{table1}, the reflection coefficient is always an even function of $k_x$. 
Hence, the approximations~\eqref{parabolic2}--\eqref{hyperbolic2} can be used to describe the reflection coefficient of the structure with an arbitrary symmetry. As for the transmission, only the symmetries 
p11m, p11g, p2mg, p2mm, and p1m1 provide the assumptions used to derive approximations~\eqref{parabolic2}--\eqref{hyperbolic2}.

The structures described by the symmetry groups p1 and p2 require the use of more general approximations. 
While the denominator for the transmission approximation will be of the same form as for the reflection [see Eq.~\eqref{Eq4}], the numerators in Eqs.~\eqref{parabolic2}--\eqref{hyperbolic2} might have additional terms in $k_x$ and $k_x \omega$.

Now let us consider two important symmetries, which allow us to simplify the general representations~\eqref{hyperbolic}~and~\eqref{parabolic}.

\subsection{Vertical plane of symmetry}
In this subsection, we will focus on hyperbolic approximation~\eqref{hyperbolic} for the structures with the $yOz$ symmetry plane (symmetry groups p2mm, p1m1, p2mg). 
Usually, the effects caused by the structure symmetry emerge when the incident wave has the same symmetry as the structure. 
Therefore, in this subsection we analyze the case of a normally incident plane wave ($k_x = 0$).

Let us study the symmetry of the field distribution of the modes with respect to the symmetry plane.
Approximation~\eqref{hyperbolic} assumes that at $k_x = 0$ the structure supports two modes with the corresponding complex frequencies $\omega_{p1}$ and $\omega_{p2}$.
One can show that the modes of a symmetric structure are either symmetric or antisymmetric~\cite{Belotelov:2014:prb}. 
Moreover, exactly one of two modes is symmetric, while the other one is antisymmetric. 
Without loss of generality, we assume that the mode with frequency $\omega_{p2}$ is the antisymmetric one.

The antisymmetric modes cannot be excited in a symmetric structure by the normally incident plane wave~\cite{Gippius:2005:prb, Belotelov:2014:prb}.
This means that the corresponding pole $\omega_{p2}$ should not affect the transmission spectrum $T(0, \omega)$.
It is possible if and only if one of the zeros compensates the pole (e.g.\ if $\omega_{z2} = \omega_{p2}$).
Taking this fact into account, we rewrite Eq.~\eqref{hyperbolic2} in the following form: 
\begin{equation}
\label{hyperbolic3}
T(k_x,\omega) = t\frac{v_g^2 k_x^2 - \gamma(\omega-\omega_{z1})(\omega-\omega_{p2})}{v_g^2 k_x^2 - (\omega-\omega_{p1})(\omega-\omega_{p2})}.
\end{equation}

There are two channels of the mode decay in PCS: the scattering into one of the diffraction orders and the ohmic losses. Due to the symmetry, the antisymmetric mode cannot scatter into the zeroth diffraction order. Hence, the antisymmetric mode of the lossless structure does not decay in time and its frequency is always real.

\subsection{Horizontal plane of symmetry}
In this subsection, we consider the structures with the $xOy$ symmetry plane (symmetry groups p2mm and p11m). 
Besides, the results of the subsection are valid as well for the structures with symmetry groups p11g and p2mg. 
We start our analysis from recalling the symmetry and energy-conservation consequences for the conventional Fano line-shape~\eqref{Fano}.
Then we use these consequences to study the $\omega$--$k_x$ Fano line-shapes~\eqref{parabolic}~and~\eqref{hyperbolic}.

Let us consider the single-resonance structure with the transmission spectrum defined by the Fano line-shape~\eqref{Fano}.
In the case of the lossless structure, the transmission and refection coefficients satisfy the energy conservation law that can be represented as $|T(\omega)|^2 + |R(\omega)|^2 = 1$.
The more general formulation of the energy conservation law requires the unitarity of the scattering matrix~\cite{Gippius:2005:prb}
\begin{equation}
\label{smatrix_sym}
S = 
\begin{bmatrix}
R&T\\
T&R
\end{bmatrix}.
\end{equation}
Note that this form of the scattering matrix takes account of the symmetry of the considered structure (see Table~\ref{table1}).

It can be shown that if both $R$ and $T$ have the form of Eq.~\eqref{Fano}, the unitarity of the scattering matrix~\eqref{smatrix_sym} implies the following relation~\cite{my:golovastikov:2013:jopt}: 
\begin{equation}
\label{unit_main}
\omega_z = \Re \omega_p \pm \mathrm{i} \frac{r}{t} \Im \omega_p,
\end{equation}
where $r$ is the non-resonant reflection coefficient. 
The plus (minus) sign corresponds to the antisymmetric (symmetric) mode with respect to the $xOy$ symmetry plane. 
Equation~\eqref{unit_main} defines the zero of the transmission coefficient $T$ [see Eq.~\eqref{Fano}]. 
The zero of the reflection coefficient $R$ is also represented by Eq.~\eqref{unit_main} with the $r$ and $t$ interchanged~\cite{my:golovastikov:2013:jopt}.

By formally replacing $\omega$ in Eq.~\eqref{Fano} with $k_x$, we obtain a $k_x$-Fano line-shape~\cite{Popov:1986:oa, Neviere:1995:josaa, Lomakin:2007:trant}.
In this case, Eq.~\eqref{unit_main} is also true~\cite{Popov:1986:oa}. 
If we replace $\omega$ in Eq.~\eqref{Fano} with $k_x^2$ we obtain the Fano line-shape that takes account of two symmetric poles in the vicinity of $k_x = 0$. 
In this case, Eq.~\eqref{unit_main} remains valid as well.

Let us use Eq.~\eqref{unit_main} to analyze the $\omega$--$k_x$ Fano line-shapes.
Equations~\eqref{parabolic}~and~\eqref{hyperbolic} can be written in the following general form: 
\begin{equation}
\label{para_hyp}
T(k_x, \omega) = t\frac{k_x^2 - Z(\omega)}{k_x^2 - P(\omega)},
\end{equation}
where $Z(\omega)$ and $P(\omega)$ are the polynomials of order two or three.

Let us fix the value of angular frequency $\omega$ and consider the transmission coefficient $T(\omega, k_x)$ as a function of $k_x^2$. 
In this case, Eq.~\eqref{para_hyp} will take the form of the conventional Fano line-shape~\eqref{Fano}. Hence, Eq.~\eqref{unit_main} can be written in the following form: 
\begin{equation}
\label{unit_main2}
Z(\omega) = \Re P(\omega) \pm \mathrm{i} \frac{r}{t} \Im P(\omega).
\end{equation}
By replacing $P(\omega)$ and $Z(\omega)$ with their Taylor expansions, we obtain the following equation:
\begin{equation}
\label{unit_main3}
z_k = \Re p_k \pm \mathrm{i} \frac{r}{t} \Im p_k.
\end{equation}
Thus, in the case of a lossless symmetric structure the coefficients in the numerator of representations~\eqref{parabolic}~and~\eqref{hyperbolic} are determined by the coefficients in the denominator.

Let us note that
for the structure with horizontal plane of symmetry the value of
 $\mathrm{i}\,r/t$ in Eq.~\eqref{unit_main3} is 
always
real~\cite{Fan:2003:josaa}, hence $z_k$ are real as well. 
Therefore, $\omega_{z1}$ and $\omega_{z2}$ in Eqs.~\eqref{hyperbolic2},~\eqref{parabolic2} are either real or complex-conjugate. 
This results in the total transmission/reflection at certain frequencies of the incident light, which is a well-known phenomenon for the Fano resonances in lossless structures~\cite{Gippius:2005:prb, Popov:1986:oa, Fan:2003:josaa}.

\section{\label{section4}Causality}

In this section, we will study the causality property of the transmission (reflection) coefficients. According to the general principle of causality, the caused effect cannot occur before the cause. In our case it means that the light can appear in the region under the structure only after the moment of time at which the light impinges the structure. Obviously, the transmission and reflection coefficients rigorously calculated from Maxwell's equations are always causal. It is well known that the Fano line-shape~\eqref{Fano} is causal if and only if $\Im \omega_p < 0$~\cite{Nussenzveig:1972, Tikhodeev:2002:prb}. In what follows we will investigate the causality conditions for the approximations~\eqref{parabolic2} and~\eqref{hyperbolic2}. We will study the causality in terms of the impulse response of the structure.

Let us represent the incident light beam through its spectrum: 
\begin{equation}
\label{incident_field}
A_{\rm inc}(x,z,t) = \frac{1}{(2\pi)^2}\iint G(k_x, \omega) \mathrm{e}^{\mathrm{i} (k_x x - k_z z - \omega t)} \, \mathrm{d} k_x \, \mathrm{d} \omega,
\end{equation}
where $k_x^2+k_z^2 = n_{\rm s}^2$, $n_{\rm s}$ is the surrounding medium refractive index. Since Eq.~\eqref{incident_field} is a plane wave expansion, we can define the transmitted field as follows:
\begin{equation}
\label{t_field}
\begin{aligned}
&A_{\rm tr}(x,z,t) \\
&= \frac{1}{(2\pi)^2}\iint T(k_x, \omega) G(k_x, \omega) \mathrm{e}^{\mathrm{i} (k_x x - k_z (z+h) - \omega t)} \, \mathrm{d} k_x \, \mathrm{d} \omega,
\end{aligned}
\end{equation}
where $T(k_x, \omega)$ is the transmission coefficient of the structure, $h$ is the thickness of the structure.

Now let us investigate the transmitted field at the structure lower interface $A_{\rm tr}(x, -h, t)$ for the case of the incident pulse corresponding to the Dirac delta function on the structure upper interface [$A_{\rm inc}(x,0,t) = \delta(x)\cdot\delta(t)$]. In this case, the incident pulse spectrum is equal to unity [$G(k_x, \omega) = 1$], and the transmitted field distribution 
\begin{equation}
\label{impresp}
\begin{aligned}
h(x,t) &= A_{\rm tr}(x, -h, t) 
\\&= \frac{1}{4\pi^2}\iint T(k_x, \omega) \mathrm{e}^{\mathrm{i} (k_x x - \omega t)} \, \mathrm{d} k_x \, \mathrm{d} \omega
\end{aligned}
\end{equation}
can be considered as the impulse response of the structure.

Let us study the impulse responses for the $\omega$--$k_x$ Fano representations~\eqref{parabolic2}~and~\eqref{hyperbolic2}.
Note that the numerators of Eqs.~\eqref{parabolic2},~\eqref{hyperbolic2} may be considered causal since the corresponding impulse responses can be expressed in terms of the delta function and its derivatives. 
In what follows we will study the impulse response that takes account of only the denominators of Eqs.~\eqref{parabolic2},~\eqref{hyperbolic2}.

For parabolic approximation~\eqref{parabolic2} the impulse response can be easily calculated. 
Assuming $\Im ( \omega_p + k_x^2/\beta ) < 0,$ or 
\begin{equation}
\label{cause_parab}
\Im \omega_p < 0, \;\; \Im\beta \geqslant 0,
\end{equation}
we obtain the following impulse response: 
\begin{equation}
\label{impresp_parab}
\begin{aligned}
h(x,t) &= \frac{1}{4\pi^2}\iint \frac{\mathrm{e}^{\mathrm{i} (k_x x - \omega t)}}{k_x^2 - \beta(\omega-\omega_p)} \, \mathrm{d} k_x \, \mathrm{d} \omega
\\&=
\begin{cases}
\frac{1}{2}\sqrt{\frac{\mathrm{i}}{\pi\beta  t}} \mathrm{e}^{-\mathrm{i} \omega_p t } \mathrm{e}^{\frac{\mathrm{i} x^2 \beta}{4t}},  & t > 0;\\
0, & t<0.\\
\end{cases}
\end{aligned}
\end{equation}

According to Eq.~\eqref{impresp_parab}, $h(x,t)$ is zero at $t<0$. 
This means that causality takes place in the non-relativistic sense, i.e.\ the light appears under the structure only after the moment in which the incident beam impinges it. 
However, Eq.~\eqref{impresp_parab} assumes that the superluminal light propagation in the transverse direction can occur.
In order to demonstrate this fact, let us consider the dispersion equation for parabolic approximation [i.e.\ equate the denominator of Eq.~\eqref{parabolic2} to zero]: 
\begin{equation}
k_x^2 = \beta(\omega-\omega_p).
\end{equation}
From this equation we can deduce the complex group velocity of the mode, $v_g = {\mathrm{d} \omega}/{\mathrm{d} k_x} = 2 k_x / \beta$. 
Its real part defining the propagation velocity of the mode~\cite{Peatross:2000:prl} can be arbitrarily large. 
In particular, it can overcome the speed of light at the large values of $k_x$. 
Hence, the parabolic approximation violates the relativistic causality condition.
The latter can be formulated as follows~\cite{Nussenzveig:1972}: $h(x,t) = 0$ if $|x|>ct$, where $c$ is the speed of light.

Let us investigate impulse response for the hyperbolic approximation~\eqref{hyperbolic2}.
By substituting the denominator of Eq.~\eqref{hyperbolic2} into Eq.~\eqref{impresp} we obtain the following:
\begin{widetext}
\begin{equation}
\label{hyperimpresp}
\begin{aligned}
h(x,t) &= \frac{1}{4\pi^2}\iint \frac{\mathrm{e}^{\mathrm{i} (k_x x - \omega t)}}{v_g^2 k_x^2 - (\omega-\omega_{p1})(\omega-\omega_{p2})} \, \mathrm{d} k_x \, \mathrm{d} \omega
\\&=
\begin{cases}
\dfrac{1}{2 v_g} \exp\left(-\mathrm{i} \dfrac{\omega_{p1} + \omega_{p2}}2 t\right) J_0\left(\dfrac{\omega_{p1} - \omega_{p2}}2\sqrt{t^2-\dfrac{x^2}{v_g^2}} \right),  & |x| < v_g t;\\
0, & |x| > v_gt,\\
\end{cases}
\end{aligned}
\end{equation}
\end{widetext}
where $J_0(x)$ is the Bessel function of the first kind of order zero. 
The detailed derivation of Eq.~\eqref{hyperimpresp} is presented in the Appendix.
In the derivation, we supposed that 
\begin{equation}
\label{cause_hyperb}
\Im \omega_{p1} < 0,\;\; \Im \omega_{p2} < 0, \;\; v_g \in \mathbb{R}.
\end{equation}
According to Eq.~\eqref{hyperimpresp}, if we further assume that $v_g \leqslant c$, the relativistic causality condition will take place.

Let us note that the dispersion relation for hyperbolic approximation, 
\begin{equation}
v_g^2 k_x^2 = (\omega-\omega_{p1})(\omega-\omega_{p2}),
\end{equation}
defines the hyperbola with asymptotes $\omega = \pm v_g k_x$.
This means that $\pm v_g$ is the group velocity of the mode at $|k_x|\gg1$. 
In other words, $v_g$ is the group velocity of the mode in the empty lattice approximation.

Thus, we have shown that the conditions~\eqref{cause_parab} and~\eqref{cause_hyperb} provide causality of the parabolic and hyperbolic approximations, respectively. 
Moreover, the hyperbolic approximation is causal in a relativistic sense.
We can call the expressions~\eqref{cause_parab},~\eqref{cause_hyperb} the causality conditions, since their violation leads to the non-causality of the corresponding approximations~\eqref{parabolic2},~\eqref{hyperbolic2}.

Let us use the causality condition~\eqref{cause_hyperb}, assuming that both $xOy$ and $yOz$ are the symmetry planes of the structure (symmetry groups p2mm and p2mg). 
In this case, Eqs.~\eqref{hyperbolic3},~\eqref{unit_main3} lead to the following elegant form of the $\omega$--$k_x$ Fano line-shape for a symmetric lossless structure: 
\begin{equation}
\label{hyperbolic4}
\begin{aligned}
T(k_x,\omega) &= t\frac{v_g^2 k_x^2 - (\omega-\omega_{zr})(\omega-\omega_{p2})}{v_g^2 k_x^2 - (\omega-\omega_{p1})(\omega-\omega_{p2})},
\\
R(k_x,\omega) &= r\frac{v_g^2 k_x^2 - (\omega-\omega_{zt})(\omega-\omega_{p2})}{v_g^2 k_x^2 - (\omega-\omega_{p1})(\omega-\omega_{p2})},
\end{aligned}
\end{equation}
where $\omega_{p1}, t, r \in \mathbb{C}$;  $v_g, \omega_{p2}, \omega_{zt}, \omega_{zr}, \mathrm{i}\frac{r}{t} \in \mathbb{R}$;
$|r|^2+|t|^2=1$;
$\omega_{zt} = \Re \omega_{p1} \pm \mathrm{i}\frac{r}{t}\Im \omega_{p1}$;
$\omega_{zr} = \Re \omega_{p1} \pm \mathrm{i}\frac{t}{r}\Im \omega_{p1}$.

\section{\label{section5}Conclusion}
We have presented an $\omega$--$k_x$ generalization of the Fano line-shape for photonic crystal slabs. 
The conventional Fano line-shape is an approximation of the transmission (reflection) spectrum as a function of either frequency or in-plane wave vector that takes account of a single resonance.
The proposed generalization describes the scattering amplitude as a function of both frequency and in-plane wave vector of the incident light. Two particular line-shapes, parabolic and hyperbolic, have been obtained. These line-shapes can be used to approximate the transmission and reflection spectrum of subwavelength photonic crystal slabs. 
The hyperbolic approximation can be used to describe guided-mode resonances where two poles (resonances) at every $k_x$ are present. The parabolic approximation can be used to approximate the transmission coefficient in a small vicinity of a single pole. This approximation is particularly useful for describing cavity or Fabry--P\'{e}rot resonances.
We have studied the consequences of reciprocity, symmetry and causality to obtain the most simple form of these approximations.

%
%

\begin{acknowledgments}
The work was funded by the Russian Science Foundation grant 14-19-00796.
\end{acknowledgments}

\appendix*
\section{\label{appendix}Impulse response for hyperbolic approximation}

Consider the following transfer function:
\begin{equation}
\begin{aligned}
T(k_x, \omega) 
&= \frac{1}{v_g^2 k_x^2 - (\omega - \omega_{p1}) (\omega - \omega_{p2})} 
\\&= \frac{-1}{ (\omega - a)^2 - \left( b^2 + v_g^2 k_x^2\right)} ,
\end{aligned}
\end{equation}
where $a = (\omega_{p1} + \omega_{p2}) / 2$, $b = (\omega_{p1} - \omega_{p2}) / 2$.
In this section, we will calculate the corresponding impulse response function [Eq.~\eqref{impresp}].
We will start with taking the integral with respect to the angular frequency $\omega$.
To do this, let us consider the following integral: 
\begin{equation}
\label{lab1}
\frac{1}{2\pi}\int_{-\infty}^{+\infty} \frac{\mathrm{e}^{-\mathrm{i} \omega t}}{(\omega-a)^2 - b^2} \, \mathrm{d} \omega 
= -\frac{\sin (bt)}{b} \mathrm{e}^{-\mathrm{i} a t} \theta(t),
\end{equation}
where $\Im(a \pm b) < 0$ and $\theta(t)$ is the Heaviside step function.
According to Eq.~\eqref{lab1}, the Fourier transform of $T(k_x, \omega)$ with respect to $\omega$ can be calculated as 
\begin{equation}
\label{lab2}
\begin{aligned}
&-\frac{1}{2\pi}\int_{-\infty}^{+\infty} \frac{\mathrm{e}^{-\mathrm{i} \omega t}}{ (\omega - a)^2 - (b^2 + v_g^2 k_x^2)} \, \mathrm{d} \omega 
\\&\;\;\;=
\mathrm{e}^{-\mathrm{i} a t} \theta(t) \cdot \frac{\sin \left(t\sqrt{b^2 + v_g^2 k_x^2}\right)}{\sqrt{b^2 + v_g^2 k_x^2}}.
\end{aligned}
\end{equation}
This derivation requires $\Im\left(a \pm \sqrt{b^2 + v_g^2 k_x^2}\right) < 0$.
The latter inequality holds true for all real $k_x$ when $\Im \omega_{p1,2} < 0$ and $v_g$ is real.

Now let us calculate the second Fourier transform with respect to $k_x$. To do this, let us use the following integral identity~\cite{Bateman:1954}:
\begin{equation}
\begin{aligned}
&\frac{1}{2\pi}\int_{-\infty}^{+\infty}\frac{\sin\left(y\sqrt{k^2+q^2}\right)}{\sqrt{k^2+q^2}} \mathrm{e}^{-\mathrm{i} k x} \, \mathrm{d} k 
\\&\;\;\;= 
\begin{cases}
\frac{1}{2} J_0\left(q\sqrt{y^2-x^2} \right),  & |x| < y;\\
0, & |x| > y.\\
\end{cases}
\end{aligned}
\end{equation}
Using this equation we find the impulse response as the Fourier transform of~\eqref{lab2} in the following form: 
\begin{equation}
\label{lab3}
\begin{aligned}
h(x,t) &= \frac{\mathrm{e}^{-\mathrm{i} a t} \theta(t)}{2\pi}
\int_{-\infty}^{+\infty} 
\frac{\sin \left(t\sqrt{b^2 + v_g^2 k_x^2}\right)}{\sqrt{b^2 + v_g^2 k_x^2}}
 \mathrm{e}^{\mathrm{i} k_x x} \, \mathrm{d} k_x
\\&=
\mathrm{e}^{-\mathrm{i} a t} \theta(t)
\begin{cases}
\frac{1}{2 v_g} J_0\left(\frac{b}{v_g}\sqrt{v_g^2 t^2-x^2} \right),  & |x| < v_g t;\\
0, & |x| > v_g t.\\
\end{cases}
\end{aligned}
\end{equation}
From the last equation it is easy to obtain Eq.~\eqref{hyperimpresp}.

\bibliographystyle{unsrt}
\bibliography{SpatioTemporal}

\begin{thebibliography}{10}

\bibitem{Beutler:1935:zp}
H.~Beutler.
\newblock {{\"U}ber Absorptionsserien von Argon, Krypton und Xenon zu Termen
  zwischen den beiden Ionisierungsgrenzen $^2$P$_3^{2/0}$ und $^2$P$_1^{2/0}$}.
\newblock {\em Zeitschrift f{\"u}r Physik}, 93(3-4):177--196, 1935.

\bibitem{Fano:1935:nc}
U.~Fano.
\newblock Sullo spettro di assorbimento dei gas nobili presso il limite dello
  spettro d’arco.
\newblock {\em Nuovo Cimento}, 12:154--161, 1935.

\bibitem{Fano:1961:pr}
U.~Fano.
\newblock Effects of configuration interaction on intensities and phase shifts.
\newblock {\em Phys. Rev.}, 124(6):1866--1878, Dec 1961.

\bibitem{Miroshnichenko:2010:rmp}
Andrey~E. Miroshnichenko, Sergej Flach, and Yuri~S. Kivshar.
\newblock {Fano resonances in nanoscale structures}.
\newblock {\em Rev. Mod. Phys.}, 82:2257--2298, Aug 2010.

\bibitem{Collin:2014:rpp}
St{\'e}phane Collin.
\newblock Nanostructure arrays in free-space: optical properties and
  applications.
\newblock {\em Reports on Progress in Physics}, 77(12):126402, 2014.

\bibitem{Zhou:2014:pqe}
Weidong Zhou, Deyin Zhao, Yi-Chen Shuai, Hongjun Yang, Santhad Chuwongin,
  Arvinder Chadha, Jung-Hun Seo, Ken~X. Wang, Victor Liu, Zhenqiang Ma, and
  Shanhui Fan.
\newblock {Progress in 2D photonic crystal Fano resonance photonics}.
\newblock {\em Prog. Quantum Electron.}, 38(1):1--74, 2014.

\bibitem{Kirilenko:1993:em}
A.~A Kirilenko and B.~G. Tysik.
\newblock Connection of s-matrix of waveguide and periodical structures with
  complex frequency spectrum.
\newblock {\em Electromagn.}, 13(3):301--318, 1993.

\bibitem{Centeno:2000:prb}
E.~Centeno and D.~Felbacq.
\newblock Optical bistability in finite-size nonlinear bidimensional photonic
  crystals doped by a microcavity.
\newblock {\em Phys. Rev. B}, 62:R7683--R7686, Sep 2000.

\bibitem{Collin:2001:prb}
S.~Collin, F.~Pardo, R.~Teissier, and J.-L. Pelouard.
\newblock Strong discontinuities in the complex photonic band structure of
  transmission metallic gratings.
\newblock {\em Phys. Rev. B}, 63:033107, Jan 2001.

\bibitem{Tikhodeev:2002:prb}
S.~G. Tikhodeev, A.~L. Yablonskii, E.~A. Muljarov, N.~A. Gippius, and Teruya
  Ishihara.
\newblock Quasiguided modes and optical properties of photonic crystal slabs.
\newblock {\em Phys. Rev. B}, 66(4):045102, Jul 2002.

\bibitem{Sarrazin:2003:prb}
Micha\"el Sarrazin, Jean-Pol Vigneron, and Jean-Marie Vigoureux.
\newblock {Role of Wood anomalies in optical properties of thin metallic films
  with a bidimensional array of subwavelength holes}.
\newblock {\em Phys. Rev. B}, 67:085415, Feb 2003.

\bibitem{Gippius:2005:prb}
N.~A. Gippius, S.~G. Tikhodeev, and T.~Ishihara.
\newblock Optical properties of photonic crystal slabs with an asymmetrical
  unit cell.
\newblock {\em Phys. Rev. B}, 72(4):045138, Jul 2005.

\bibitem{Lomakin:2006:trant}
Vitaliy Lomakin and E.~Michielssen.
\newblock Transmission of transient plane waves through perfect electrically
  conducting plates perforated by periodic arrays of subwavelength holes.
\newblock {\em IEEE Trans. Antennas Propag.}, 54(3):970--984, March 2006.

\bibitem{my:bykov:2013:jlt}
D.~A. Bykov and L.~L. Doskolovich.
\newblock {Numerical methods for calculating poles of the scattering matrix
  with applications in grating theory}.
\newblock {\em J. Lightw. Technol.}, 31(5):793--801, March 2013.

\bibitem{Popov:1986:oa}
E.~Popov, L.~Mashev, and D.~Maystre.
\newblock Theoretical study of the anomalies of coated dielectric gratings.
\newblock {\em Opt. Acta}, 33(5):607--619, 1986.

\bibitem{Neviere:1995:josaa}
M.~Nevi\`{e}re, E.~Popov, and R.~Reinisch.
\newblock Electromagnetic resonances in linear and nonlinear optics:
  phenomenological study of grating behavior through the poles and zeros of the
  scattering operator.
\newblock {\em J. Opt. Soc. Am. A}, 12(3):513--523, 1995.

\bibitem{Lomakin:2007:trant}
Vitaliy Lomakin and E.~Michielssen.
\newblock Beam transmission through periodic subwavelength hole structures.
\newblock {\em IEEE Trans. Antennas Propag.}, 55(6):1564--1581, June 2007.

\bibitem{Shipman:2005:pre}
Stephen~P. Shipman and Stephanos Venakides.
\newblock Resonant transmission near nonrobust periodic slab modes.
\newblock {\em Phys. Rev. E}, 71:026611, Feb 2005.

\bibitem{Shipman:2013:jmp}
Stephen~P. Shipman and Aaron~T. Welters.
\newblock Resonant electromagnetic scattering in anisotropic layered media.
\newblock {\em J. Math. Phys.}, 54(10):103511, 2013.

\bibitem{Lalanne:2008:nat}
Haitao Liu and Philippe Lalanne.
\newblock Microscopic theory of the extraordinary optical transmission.
\newblock {\em Nature}, 452:728--731, April 2008.

\bibitem{Fan:2003:josaa}
Shanhui Fan, Wonjoo Suh, and J.~D. Joannopoulos.
\newblock {Temporal coupled-mode theory for the Fano resonance in optical
  resonators}.
\newblock {\em J. Opt. Soc. Am. A}, 20(3):569--572, 2003.

\bibitem{Wood:1902:pm}
R.W. Wood.
\newblock Xlii. on a remarkable case of uneven distribution of light in a
  diffraction grating spectrum.
\newblock {\em Philos. Mag. Ser. 6}, 4(21):396--402, 1902.

\bibitem{Hessel:1965:ao}
A.~Hessel and A.~A. Oliner.
\newblock {A new theory of Wood's anomalies on optical gratings}.
\newblock {\em Appl. Opt.}, 4(10):1275--1297, Oct 1965.

\bibitem{Wang:1993:ao}
S.~S. Wang and R.~Magnusson.
\newblock Theory and applications of guided-mode resonance filters.
\newblock {\em Appl. Opt.}, 32(14):2606--2613, May 1993.

\bibitem{Mutlu:2012:ol}
Mehmet Mutlu, Ahmet~E. Akosman, and Ekmel Ozbay.
\newblock Broadband circular polarizer based on high-contrast gratings.
\newblock {\em Opt. Lett.}, 37(11):2094--2096, Jun 2012.

\bibitem{my:golovastikov:2013:jopt}
D.~A. Bykov, L.~L. Doskolovich, N.~V. Golovastikov, and V.~A. Soifer.
\newblock {Time-domain differentiation of optical pulses in reflection and in
  transmission using the same resonant grating}.
\newblock {\em J. Opt.}, 15:105703, 2013.

\bibitem{Belotelov:2014:prb}
V.~I. Belotelov, L.~E. Kreilkamp, A.~N. Kalish, I.~A. Akimov, D.~A. Bykov,
  S.~Kasture, V.~J. Yallapragada, Achanta~Venu Gopal, A.~M. Grishin, S.~I.
  Khartsev, M.~Nur-E-Alam, M.~Vasiliev, L.~L. Doskolovich, D.~R. Yakovlev,
  K.~Alameh, A.~K. Zvezdin, and M.~Bayer.
\newblock Magnetophotonic intensity effects in hybrid metal-dielectric
  structures.
\newblock {\em Phys. Rev. B}, 89:045118, Jan 2014.

\bibitem{Liu:2001:prb}
H.~Liu, G.~X. Li, K.~F. Li, S.~M. Chen, S.~N. Zhu, C.~T. Chan, and K.~W. Cheah.
\newblock {Linear and nonlinear Fano resonance on two-dimensional magnetic
  metamaterials}.
\newblock {\em Phys. Rev. B}, 84:235437, Dec 2011.

\bibitem{Moharam:1995:josaa}
M.~G. Moharam, Eric~B. Grann, Drew~A. Pommet, and T.~K. Gaylord.
\newblock Formulation for stable and efficient implementation of the rigorous
  coupled-wave analysis of binary gratings.
\newblock {\em J. Opt. Soc. Am. A}, 12(5):1068--1076, 1995.

\bibitem{Li:1996:josaa}
Lifeng Li.
\newblock Formulation and comparison of two recursive matrix algorithms for
  modeling layered diffraction gratings.
\newblock {\em J. Opt. Soc. Am. A}, 13(5):1024--1035, May 1996.

\bibitem{Christ:2003:prl}
A.~Christ, S.~G. Tikhodeev, N.~A. Gippius, J.~Kuhl, and H.~Giessen.
\newblock Waveguide-plasmon polaritons: Strong coupling of photonic and
  electronic resonances in a metallic photonic crystal slab.
\newblock {\em Phys. Rev. Lett.}, 91:183901, Oct 2003.

\bibitem{Yablonskii:2002:pssa}
A.L. Yablonskii, E.A. Muljarov, N.A. Gippius, S.G. Tikhodeev, and T.~Ishihara.
\newblock Optical properties of polaritonic crystal slab.
\newblock {\em physica status solidi (a)}, 190(2):413--419, 2002.

\bibitem{Scheidemann:2005}
Volker Scheidemann.
\newblock {\em Introduction to complex analysis in several variables}.
\newblock Birkh{\"a}user Verlag Basel, Boston, Berlin, 2005.

\bibitem{Kopsky:2002}
V.~Kopsk\'{y} and D.~B. Litvin.
\newblock {\em International tables for crystallography. Volume E: Subperiodic
  Groups}.
\newblock Kluwer academic publishers Dordrecht, Boston, London, 2002.

\bibitem{Nussenzveig:1972}
Herch~Moyses Nussenzveig.
\newblock {\em Causality and dispersion relations}.
\newblock Academic Press, New York, 1972.

\bibitem{Peatross:2000:prl}
J.~Peatross, S.~A. Glasgow, and M.~Ware.
\newblock Average energy flow of optical pulses in dispersive media.
\newblock {\em Phys. Rev. Lett.}, 84:2370--2373, Mar 2000.

\bibitem{Bateman:1954}
Arthur Erd{\'e}lyi and Harry Bateman.
\newblock {\em Tables of integral transforms, Vol I}.
\newblock McGraw-Hill New York, 1954.

\end{thebibliography}

\end{document}